\documentclass[letterpaper]{article} 
\usepackage{aaai24}  
\usepackage{times}  
\usepackage{helvet}  
\usepackage{courier}  
\usepackage[hyphens]{url}  
\usepackage{graphicx} 
\usepackage{natbib}  
\usepackage{caption} 
\usepackage{algorithm}
\usepackage{algorithmic}
\usepackage{afterpage}
\nocopyright
\usepackage{color}

\usepackage{amsmath}
\usepackage{amsfonts}       
\usepackage{dsfont}
\usepackage{booktabs, multirow, tabularx, threeparttable}
\usepackage{subcaption}
\usepackage{comment}

\newcommand{\inlinemath}[1]{{\begin{math}#1\end{math}}}

\newcommand{\bricc}{{\sc Bricc}} 
\usepackage{newfloat}
\usepackage{listings}
\DeclareCaptionStyle{ruled}{labelfont=normalfont,labelsep=colon,strut=off} 
\floatstyle{ruled}
\newfloat{listing}{tb}{lst}{}
\floatname{listing}{Listing}
\title{Reducing Biases towards Minoritized Populations in Medical Curricular Content via Artificial Intelligence for Fairer Health Outcomes}

\author{
    Chiman Salavati\textsuperscript{\rm 1},
    Shannon Song\textsuperscript{\rm 2},
    Willmar Sosa Diaz\textsuperscript{\rm 1},
    Scott A.\ Hale\textsuperscript{\rm 3},
    Roberto E.\ Montenegro\textsuperscript{\rm 4},
    Fabricio Murai\textsuperscript{\rm 2},
    Shiri Dori-Hacohen\textsuperscript{\rm 1}
}
\affiliations{
    \textsuperscript{\rm 1}University of Connecticut\\
    \textsuperscript{\rm 2}Worcester Polytechnic Institute\\
    \textsuperscript{\rm 3}University of Oxford\\
    \textsuperscript{\rm 4}Seattle Children's Hospital
}

\begin{document}

\maketitle

\begin{abstract}
Biased information (recently termed \textit{bisinformation}) continues to be taught in medical curricula, often long after having been debunked. In this paper, we introduce \bricc{}, a first-in-class initiative that seeks to mitigate medical bisinformation using machine learning to systematically identify and flag text with potential biases, for subsequent review in an expert-in-the-loop fashion, thus greatly accelerating an otherwise labor-intensive process. A gold-standard \bricc{} dataset was developed throughout several years, and contains over 12K pages of instructional materials. Medical experts meticulously annotated these documents for bias according to comprehensive coding guidelines, emphasizing gender, sex, age, geography, ethnicity, and race. Using this labeled dataset, 
we trained, validated, and tested medical bias classifiers. We test three classifier approaches: a binary type-specific classifier, a general bias classifier; an ensemble combining bias type-specific classifiers independently-trained; and a multi-task learning (MTL) model tasked with predicting both general and type-specific biases. While MTL led to some improvement on race bias detection in terms of F1-score, it did not outperform binary classifiers trained specifically on each task. 
On general bias detection, the binary classifier achieves up to 0.923 of AUC, a 27.8\% improvement over the baseline. 
This work lays the foundations for debiasing medical curricula by exploring a novel dataset and evaluating different training model strategies. Hence, it offers new pathways for more nuanced and effective mitigation of bisinformation.
\end{abstract}

\section{Introduction}

The field of medicine is marred by a long, painful, and deleterious history of overt and covert forms of social injustice, bias, and racism, as illustrated by the American Medical Association’s recent pledge to take action to confront systemic racism~\cite{alma9977028023601701,madara2020america}. Studies continue to demonstrate that physicians possess implicit biases in a number of different areas such as race/ethnicity, gender, sex, age, weight, substance use and mental illness~\cite{fitzgerald2017implicit}. Bias in medicine leads to harmful public health costs, which are borne disproportionately by women, the economically disadvantaged, and other minoritized~\cite{wingrove2021not} groups. Harmful biases are widespread among clinicians, such as assuming psychogenic causes for physical symptoms for women, minimizing pain in people of color, or other well-documented and entrenched biases in medicine~\cite{norman2018ask}. This medical biased information (recently termed \textit{bisinformation},~\citealt{dorihacohen2021fairness}) among clinicians persists with pernicious effects on health inequities despite refuting evidence. 

Unfortunately, a main vector of transmission for bisinformation originates from materials that continue to be taught in medical schools, even long after being debunked by research~\cite{blain2024bias}. Despite numerous calls to action to deracialize and debias medical curricula and assessment content, most medical institutions continue to teach biased medicine in the preclinical years \cite{tsai2016race,rodney2016decolonization,halman2017using}. Many educators, for example, continue to inappropriately use race as a proxy for genetics or ancestry, or even as a ``risk factor'' for numerous health outcomes often erroneously associated with race (e.g., Salt Gene Hypothesis) while ignoring social or structural determinants of health (SSDoH), such as systemic racism or income inequities \cite{ali2011use,hunt2013genes,acquaviva2010perspective,karani2017commentary,metzl2014structural}. Likewise, the inappropriate use of gender and sex terms perpetuates the idea that sex and gender are binary or stagnant (vs. fluid), which can potentially alienate gender-nonconforming students and patients alike. 

By equating social identifiers to biology without social or structural context, 
medical educators are unknowingly perpetuating a curriculum that medicalizes social identities like race or gender;  reifies the false conceptualization that race and gender are a biological reality rather than social constructs; and perpetuates biased knowledge and inappropriate language use. 
Bias reduction in curricular and assessment content is key for educating future physicians in evidence-based medicine \cite{le2020first,ripp2017race}. 

\begin{table*}[t]
\renewcommand{\arraystretch}{1.2}
\centering
\footnotesize
\begin{tabular}{p{0.08\linewidth}p{0.42\linewidth}p{0.42\linewidth}}
    \toprule
    \textbf{Bias Type} & \textbf{Quote (biased or potentially biased)} & \textbf{Annotator Comment} \\
    \midrule
    Gender & \textit{Often, significant changes in a child’s growth reflect significant events in the family unit such as a mother going to work, parents separating, moving to a new home or a significant family illness} & \textit{This statement reinforces traditional family structures which stigmatizes mothers going to work, or families without a mother or two mothers. Consider omitting gendered clause.} \\
    Sex & \textit{most common in adolescent females with BMI \inlinemath{>} 30, often treated with acetazolamide and repeated therapeutic lumbar punctures, and a weight loss program.} & \textit{Consider addressing why weight and sex are relevant factors here - does being female predispose a [patient] to idiopathic intracranial hypertension, and if so, providing a citation prevents interpretation as bias.} \\
    Race & \textit{Although the incidence is lower in patients of color, the morbidity/mortality can often be higher} & \textit{No source for claim. Consider explaining more beyond ``patients of color'' as this may come across as grouping every minority group into one and making a generalized statement} \\    
    Ethnicity & \textit{While  the components of genetic versus environmental risk have not been fully established, note  the increased incidence of colorectal cancer in the Alaskan native population.} & \textit{Include citation, stratify why Alaskan Native Population is disproportionately affected} \\
    Age & \textit{New onset solid food dysphagia in anyone over 40 especially with a long history of heartburn should be considered to be esophageal cancer until proven otherwise.} & \textit{[I]s this a medical fact regarding this age group?} \\
    Geography & \textit{Had any other members of his travel group suffered the same symptoms, either  in Brazil or after returning?} & \textit{No significance of Brazil indicated in the case study} \\
    \bottomrule
\end{tabular}
\caption{Sample biased quotes from medical educational content for various types of biases, along with the annotator comment on how to make them less biased or back up identity-specific factors with citations. Note that many quotes were labeled for two or more types of bias. 
}
\label{tab:Biased_Samples}
\end{table*}

\begin{table*}[t]
\renewcommand{\arraystretch}{1.2}
\centering
\footnotesize
\begin{tabular}{p{0.12\linewidth}p{0.66\linewidth}p{0.13\linewidth}}
    \toprule
    \textbf{Bias Type} & \textbf{Quote (non-biased)} & \textbf{Type of Negative} \\
    \midrule
    Sex & \textit{In 1971, Raisman and Field reported that female rats have more dendritic spines in the preoptic area of the hypothalamus (POA) than do males.} & Explicit (EN) \\
    Age & \textit{In general, trends attributed to colorectal cancer screening in patients \inlinemath{>} 50 although  potential impact from changes in modifiable risk factors.} & Explicit (EN) \\
    Geography & \textit{Oral and oropharyngeal cancer is diagnosed each year in 40,000 Americans and kills 8,000  of them.} & Implicit (IN) \\
    Sex & \textit{The idea that some adult behavior is influenced by the sex-steroidal milieu during development had its origins  in a classic 1959 study by Phoenix and co-workers, who showed that in male guinea pigs, testosterone acts during a narrow window of time in fetal development to permanently `organize' the brain's ability to express stereotypic sexual behavior in adulthood} & Implicit (IN) \\
    Race & \textit{[..] ED physicians generally prescribe fewer opioids to African Americans regardless of clinical disorder and presentation than to non-Hispanic whites [citation]. [..]} & Extracted (XN) \\ 
    Gender & \textit{Physical exam shows a woman in moderate distress with mild jaundice and a fever of 39°c.} & Extracted (XN) \\
    Inappropriate Use of Language & \textit{His actions were impulsive with little regard for consequences with some reports suggesting he became an alcoholic and drifter.} &  Remaining (RN) \\
    \bottomrule
\end{tabular}
\caption{Sample non-biased quotes for various bias types. \small The non-biased quotes are further divided into four types of negatives: explicit, implicit, remaining, and extracted, which will be described in Section~\ref{sec:negativeTypes}. The final sample is an example of a sentence labeled for `inappropriate use of language,' but was not labeled for bias; this code is common in the Remaining Negatives group.}
\label{tab:neg_sample}
\end{table*}


Despite the urgent need for debiasing curricular content, there are several reasons why institutions continue to struggle with this issue \cite{krishnan2019addressing}: (i) faculty educators may have a significant knowledge gap; (ii) faculty educators of all backgrounds may resist confronting their own implicit biases and privilege~\cite{frey2020white}; (iii) examining educational content for bias and language misuse often entails a manual review of a cross-sectional sample \cite{tsai2016race, martin2016equitable}, leading to high variation in assessing for bias and likely results in continued bias, since not all materials are or can be examined.

Addressing and dismantling these entrenched prejudices is paramount for advancing equitable healthcare. Thus, the objective of this research is twofold. Firstly, we undertake an empirical investigation, drawing on the expertise of trained medical professionals to systematically gather, mark, and annotate instances of bias within medical texts. This process resulted in a robust and reliable dataset that reflects the multi-faceted nature of bias within medical education. Secondly, we harness this new dataset to train artificial intelligence (AI) models for discerning bias in medical text. This approach aligns with the recently introduced \textit{Fairness via AI} framework \cite{dorihacohen2021fairness}, in which AI is used carefully and deliberately to support equitable outcomes in society. Our contributions are as follows:

\begin{itemize}
    \item We curate a comprehensive labeled dataset from medical curricula amounting to two years’ worth of instructional content (e.g. lecture notes, PowerPoint slides, articles, textbooks) 
    comprising more than 4,000 quotes annotated with codes covering a wide spectrum of biases [citation redacted for anonymity]. Quotes include positive (bias) and hard negative examples (non-bias) that were reviewed by experts before they could be labeled as `bias' or `non-bias'. 
    \item We map those codes to labels of interest through a multi-stage preprocessing procedure guided by our experts'. 
    We then identify the most prevalent types of biases---those related to gender, sex, race, ethnicity, age, and geographic location---to be considered in our study.      
    \item To augment the set of `non-bias' examples, we use lexicons of social identifiers associated with each bias type to extract additional 4,391 quotes from the medical curricular content. Filtering with those identifiers ensures that no trivial samples are included. 

    \item We implement and evaluate four approaches for building a classifier to perform two major tasks include detecting bias from non-bias and detection the type of bias: (i) a type-specific classifier, (ii) a general bias classifier, (iii) an ensemble of bias type-specific classifiers, and (iv) a multi-task classifier that predicts both general and type-specific biases.
    \item We provide a thorough comparison of the models with respect to accuracy, precision, recall, F1-score, F2-score and area under ROC curve (AUC). Additionally, we contrast the performance of a Transformer model (DistilBERT) with machine learning models trained on static textual embeddings (FastText).  
    \item We develop a first-in-class bias detection classifier by fine-tuning a pre-trained DistilBERT model on our curated dataset, achieving 0.923 of AUC at detecting (general) bias within medical curricula.
\end{itemize}

By intersecting rigorous data curation with advanced computational techniques, this work illuminates sources of bias in medical education and provides a scalable solution to mitigate their influence on future generations of healthcare professionals.

\section{Related Work}


Artificial intelligence (AI) models have an immense potential for enhancing healthcare by assisting in diagnostics and treatment decision-making through objective analysis of medical records and integration with clinical decision support tools. Yet there is growing concern among researchers about the risks and ethical implications of applying AI in healthcare without proper governance \cite{althubaiti2016information,gianfrancesco2018potential,nelson2019bias}. A major cause for this concern is that current data collection practices used for gathering training data can ultimately lead to models that will produce biased outputs.



In this context, the term ``bias'' can refer to a variety of different meanings
---statistical biases, psychological biases, and societal biases---all of which raise questions over the validity and reliability of some medical research. \citet{althubaiti2016information} examines the first two types of biases in the context of epidemiological and medical research, whereas others~\cite{gianfrancesco2018potential,nelson2019bias,mittermaier2023bias} discuss how issues with data quality used for training models may favor certain populations in detriment of others.

Some proposals for reducing biases in medical research advocate for changes in medical education~\cite{cavallo2017addressing,stanciu2017ii}. \citet{althubaiti2016information} suggests that bias awareness should be integrated into medical education early on and stress the importance of transparency in reporting research findings. \citet{stone2011non} address the problem of non-conscious racial and ethnic bias in healthcare, proposing a workshop-based intervention for health care professionals.



Research in trustworthy and explainable AI also aims to create more reliable, fair and equitable models for healthcare in trustworthy and explainable AI. By carefully integrating quality, bias risk, and data fusion, it is possible to train more dependable models, while empowering them with the capability of providing explanations along with predictions, which can help to identify and mitigate residual biases~\cite{albahri2023systematic}. In a similar vein, \citet{kiyasseh2023human} has developed a strategy which helps surgical AI systems avoid algorithmic bias in assessing surgical skills across diverse hospital settings and across surgeon subcohorts.



Additionally, advocacy for formal AI governance arises to ensure responsible use and accountability, alongside potential federal legislation for assessing algorithmic bias risks. \citet{nelson2019bias} argues that, in order to prevent bias and misuse, clinicians must play a critical role in overseeing AI algorithms and validating their use in healthcare, whereas \citet{kiyasseh2023human} calls for the need of explainable models so that regulatory bodies, such as the FDA, can provide and validate frameworks to manage these biases effectively.




Complementary to the previous ones, \citet{dorihacohen2021fairness} present a multifaceted approach towards establishing fairness in AI applications by merging insights from medical education, sociology, and antiracism, as part of a new framework. They introduce ``bisinformation'' as a new concept distinct from misinformation and call for research into its nature and mitigation. They advocate for the use of AI to identify and rectify biased or harmful information that adversely affects minority groups. This is the approach we adopt in the present paper, in which we demonstrate how AI models can be used for detecting biases in medical text, as formally defined in the next section.

\begin{table}[htbp]
\centering
\caption{BRICC Dataset characteristics.}
\label{tab:data_charachteristics}
\begin{tabular}{lc}
    \toprule
    \textbf{} & \textbf{Counts} \\
    \midrule
    Number of PDF Files & 509 \\
    Total Number of Pages & 12,647 \\
    Annotated Excerpts & 4,105 \\
    Annotated Positives & 1,302 \\
    Annotated Negatives & 2,989\\
    Extracted Negatives & 4,391 \\
    \bottomrule
\end{tabular}
\end{table}

\begin{figure}[ht] 
    \centering
    \includegraphics[width=\columnwidth]{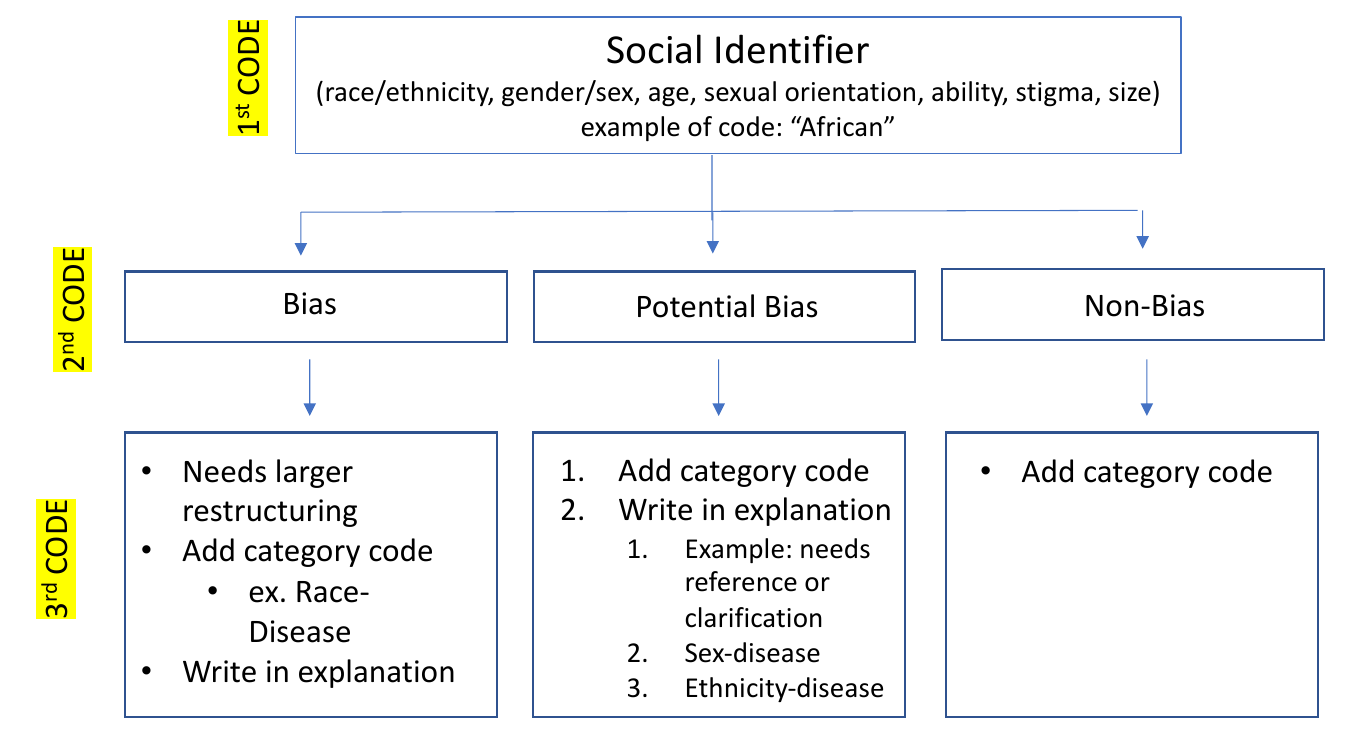} 
    \caption{\bricc{} Coding/Labeling Procedure. 
    If a social identifier is present in an excerpt, annotators coded the sample in 3 levels. 1$^\mathrm{st}$ code: tokens denoting social identifier (e.g. ``African American''); 2$^\mathrm{nd}$ code: bias label (``bias'', ``potential bias'' or ``non-bias''); and, if applicable, 3$^\mathrm{rd}$ code: identity types  (e.g., race, gender) that are relevant to the bias, along with an explanation for biased quotes.}
  \label{fig:bricc}
\end{figure}

\section{Problem Definition}
\label{sec:problemDef}

Motivated by the mandate to debias the medical curriculum at a large medical school\footnote{Institution name redacted for anonymity.}, our goal is to review medical educational content in an automated fashion and flag sentences with biased or potentially biased content, which will then be provided to experts for careful review\footnote{This ``expert-in-the-loop'' approach means that we have a preference for higher recall, and are willing to tolerate some reduced precision in order to cast a wide net. False positives will subsequently be corrected by expert review at a later stage.}.

We will define the problem formally as follows. Let \inlinemath{x} be an educational medical text excerpt, which can contain a claim, case report, incidence statistics, or any other textual content. Let \inlinemath{t} be a social identity which can be the target of bias, such as \textit{``gender,'' ``race,''} etc. Now, let \inlinemath{bias(x, t) \in \{true, false\}} be a binary variable denoting whether excerpt \inlinemath{x} exhibits bias with respect to a social identity \inlinemath{t}, or not. 

As one example, consider an excerpt in a skin cancer module that reads: \textit{``Although the incidence is lower in patients of color, the morbidity/mortality can often be higher''}. This excerpt was labeled by our expert annotators (see Section~\ref{sec:datacollection} for more details) as \textbf{biased} with respect to \textbf{race}, but not with respect to other identities (gender, geography, etc.). The annotator comment indicates: \textit{``No source for claim. Consider explaining more beyond `patients of color' as this may come across as grouping every minority group into one and making a generalized statement.''} 

By extension, we can also define \inlinemath{bias(x, \mathcal{T}) \in \{true, false\}} with respect to a set of social identities \inlinemath{\mathcal{T} = \{t_1, t_2 .. t_m\}} in the following manner: 
\begin{equation*}
    bias(x, \mathcal{T}) = true \iff \exists t \in \mathcal{T} \text{ s.t. } bias(x, t) = true
\end{equation*}

Given these definition, we now define a set of \textbf{bias classification tasks} as follows:

\begin{itemize}
    \item \textbf{Type-specific Bias Classification:} Given a social identity type \inlinemath{t} and a text excerpt \inlinemath{x}, construct a classifier that produces a prediction for \inlinemath{bias(x, t)}.
    \item \textbf{General Bias Classification:} Given a set of social identities \inlinemath{\mathcal{T}} and a text excerpt \inlinemath{x}, construct a classifier that produces a set of predictions \inlinemath{bias(x, \mathcal{T})}.
\end{itemize}

Therefore, given a set of \inlinemath{\mathcal{T} = \{t_1, t_2,\ldots, t_m\}}, we wish to construct \inlinemath{m+1} classifiers (\inlinemath{m} type-specific classifiers and one general one).


\section{Dataset}

\subsection{Strategies for Building Training Sets}

\begin{figure*}[t]
    \centering \includegraphics[width=1\linewidth]{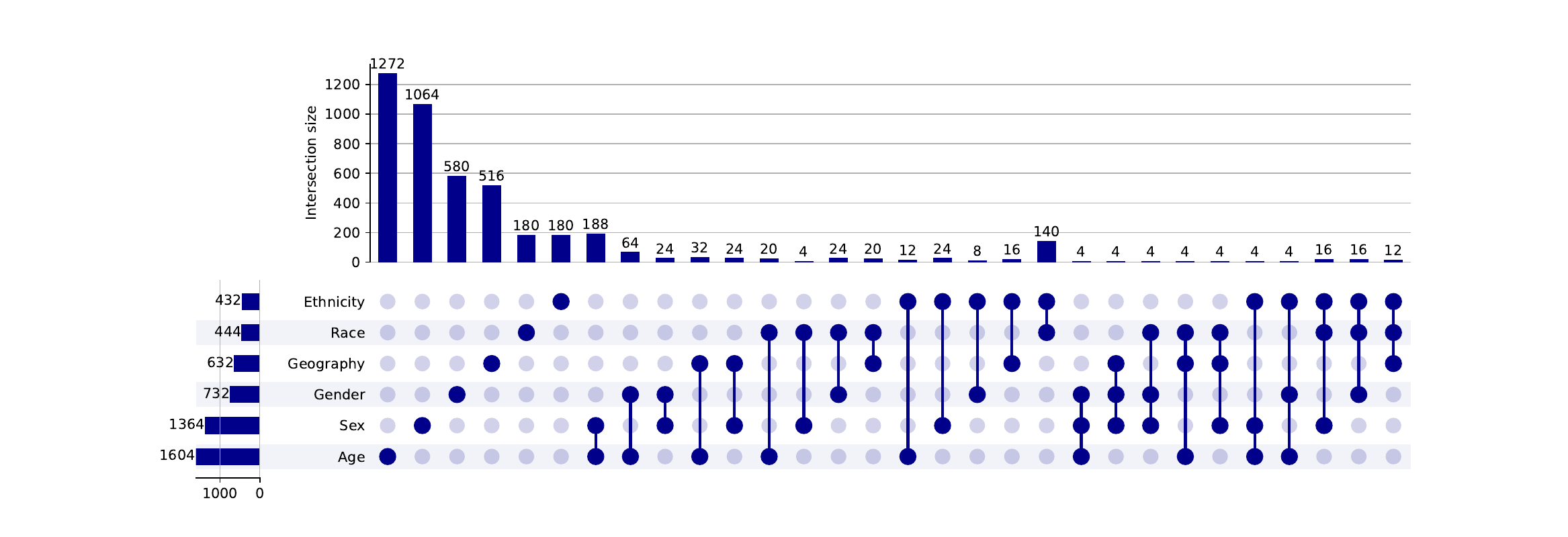}
    \caption{An Upset plot detailing the intersection between the biased quotes. A filled in circle indicates the inclusion of a specific bias type in the intersection size.}
    \label{fig:upset}
\end{figure*}

In this section we describe process of dataset collection and annotation, data extraction and pre-processing.

\subsection{Data Collection and Annotation}
\label{sec:datacollection}


Our team of medical experts has collected two years of instructional material from medical curriculum, totaling approximately 500 documents and over 12,000 pages, including both text and graphical content. Table~\ref{tab:data_charachteristics} shows a summary of the dataset statistics.
Using the ATLAS.ti software, the team has marked and subsequently coded (annotated) over 4,000 textual instances from the data. The annotators, who were at least at the pre-medical student level, underwent rigorous training. This training involved a detailed coding manual that extends beyond 20 pages and is rooted in established Content Analysis techniques. 
In what follows, we provide an overview of the annotation scheme followed by the team, which is illustrated in
Fig.~\ref{fig:bricc}. A more thorough explanation taken from the manual is reproduced in the Appendix (Fig~\ref{fig:guideline}).

The \textbf{1$^\mathrm{st}$-level code} highlights social identifiers present in the claim. The \textbf{2$^\mathrm{nd}$-level code} categorizes the claim for the presence of (potential) bias. Another dimension of coding indicates when inadequate language was used to describe social identities, such as outdated or offensive language. We do not utilize this labeling here, as we focus exclusively on bias detection. 
We then partition the data into three distinct categories---`Bias', `Potential Bias' and `Non-bias'. 
The following guidelines were used:
\begin{itemize}
  \item \textbf{Bias:} Flagging the use of stereotypes, theories of inherent group difference and advocacy of differential medical treatment based on social identities. A statement with this code definitively needs significant restructuring at minimum to become non-biased.
  \item \textbf{Potential Bias:} Flagging a statement that would be non-biased, if clearly cited, appropriate data exists and provides sufficient factual basis for this claim.
  \item \textbf{Non-bias:} Use of social identifiers in a manner that falls under the previous categories. If assertions are based on social identities, they are based on clearly cited and medically sound, compelling evidence.
  \end{itemize}
  
In some cases, the claim was coded with a \textbf{`Review'} code, signaling that the annotator was unsure of this more challenging example, and preferred to have a senior attending physician review their work.


For each biased or potentially biased excerpt in the 2$^\mathrm{nd}$-level, a \textbf{3$^\mathrm{rd}$-level code} indicates one or more social identifier category codes (e.g., ``race'') indicating which type of identity was discussed (whether in a biased or non-biased way). 
Our dataset included 17 different types of bias, but some only had a handful of examples in the entire corpus. Therefore, we focus on the six most frequent categories in our data: sex, gender, race, ethnicity, age, and geographical location (denoted as \textit{geography} for brevity).

Our panel of medical experts outlined the recommended use of these social identifiers in curricular content: 
\begin{itemize}
 \item \textbf{Sex:} sex terms such as ``female, male, AMAB, AFAB'' are acceptable when referring to population biology and are often more appropriately referenced by anatomy and genetics. Anatomic or phenotypic terms should be used with a clear purpose and are preferred (e.g., "People with ovaries"). Information about chromosomes, sex assigned at birth, and identity should be added as needed to support respectful descriptions, inclusion, and clinical reasoning.
 \item \textbf{Gender:} captures social identity such as "man, woman, boy, and girl" for individuals. Use the patient’s personally articulated gender term. Do not use gender terms when discussing population trends or outcomes-- sex terms are more appropriate and specific for these descriptions. 
  \item \textbf{Ethnicity:} inappropriate use of ethnicity (perceived shared culture) for race (perceived shared ancestry-externally imposed identity)
  \item \textbf{Race:} language that mistakes race (perceived shared ancestry- externally imposed identity) for ethnicity (perceived shared culture).
  \item \textbf{Age:} use numerical ages in describing individuals, although in some cases a descriptive term such as neonate or pediatric might be used, as is common practice in a given discipline. 
  \item \textbf{Geography:} refers to the disproportionate representation or emphasis on diseases, conditions, treatments, and health practices that are prevalent in specific geographic regions, to the potential exclusion or marginalization of those that are more common in other regions. 
  \end{itemize}

In Figure \ref{fig:upset}, we can compare the intersection between multiple bias types in our dataset. We found that the largest intersection occurs between bias types Sex and Age, Ethnicity and Race, and Gender and Age.

\subsection{Data Extraction and Pre-processing}

The ATLAS.ti format of the \bricc{} datasaet is optimized for human readability and editing - not for automated processing; therefore, we exported the data into machine readable formats, including XML, Excel spreadsheet and raw text. Once processed, the dataset contained the annotated quotes, their source (document name and page number), assigned codes, expert commentary and the identifiers for the respective medical experts. 

While the team provided meticulously detailed information at different timestamps across multiple months, we focused exclusively on the ``bias'', ``potential bias'' and ``non-bias'' categories, which relate diseases to a social identifier, while setting aside the challenge of detecting the ``inappropriate use of language'' category.

In the future work we use the additional rich source of information downstream in order to assess the relative difficulty of annotating certain quotes over others, based on the premise that quotes that required lengthy consensus-building are more ambiguous and harder to assess than those on which annotators agreed upon fairly quickly.

Let $x$ be a text excerpt containing a medical claim. 
Based on 2$^\mathrm{nd}$-level codes, dimension (ii), we define a \textbf{general bias} label as
\begin{equation}
y_\textrm{any} = \begin{cases}
1 & \textrm{if code $\in$ \{`\textit{bias}',`\textit{potential bias}',`\textit{review}'\}, } \\
0 & \textrm{otherwise}.
\end{cases}
\end{equation}

\begin{figure*}[htb]
    \centering \includegraphics[width=0.85\linewidth]{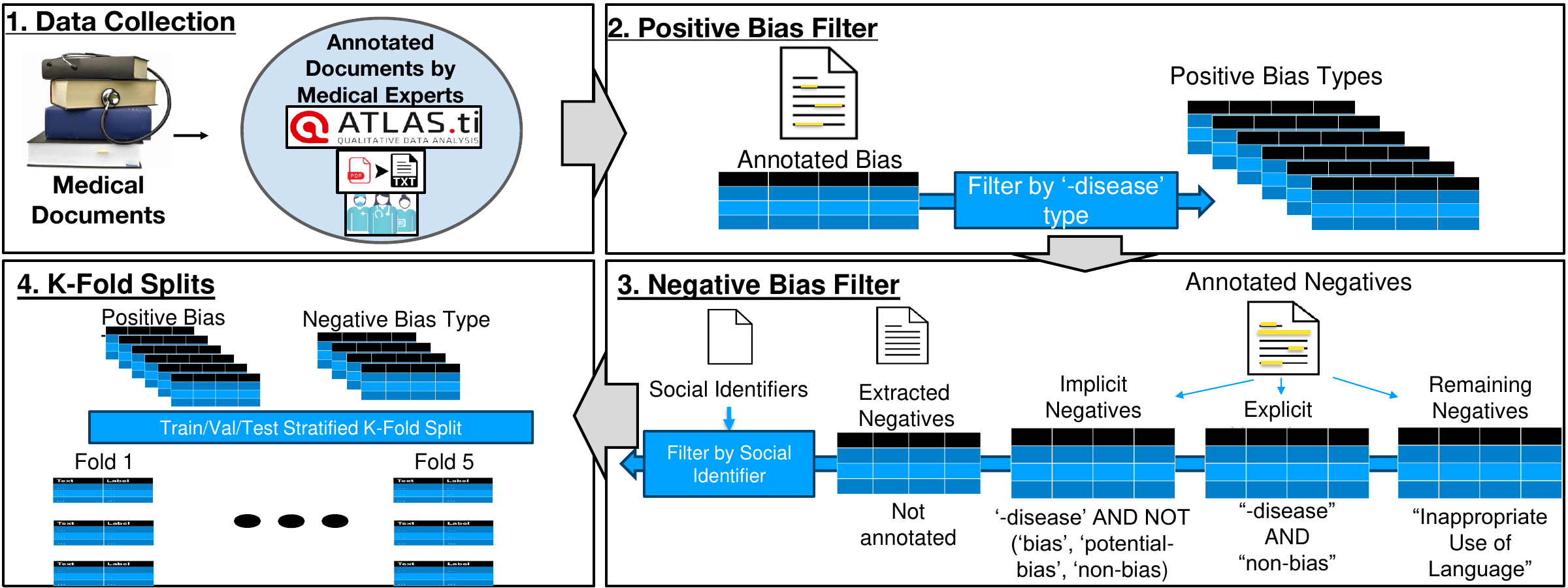}
    \caption{An overview of the proposed method from Data annotating by experts to the data splits. In part 1, we collected an annotated corpus from a team of medical experts and consolidated this to one file. In part 2, we filter the positive bias types by their respective type specific -'disease'. In part 3, we filter the negative bias subsets by different labeled conditions as well as by social identifiers for respective positive bias. In part 4, we split our data into training, validation, and testing sets by using a K-Fold split.}
    \label{fig:pipeline_a}
\end{figure*}

We opt to combine `bias', `potential bias' and `review' categories because the machine learning model is intended to be used as the first-stage of an expert-in-the-loop system. Hence the model's goal is to flag (with high recall rates) sentences to be reviewed by an expert in the second stage.

Based on 3$^\mathrm{nd}$-level codes associated with dimension (ii), we define \textbf{bias type-specific labels} for each type $t$ in \{\textit{sex, gender, race, ethnicity, age, geography}\}-disease: 
\begin{equation}
y_t = \begin{cases}
1 & \textrm{if $y_\textrm{any} = 1$ and code = $t$, } \\
0 & \textrm{otherwise}.
\end{cases}
\end{equation}

Figure~\ref{fig:pipeline_a} illustrates the three key stages of our data preprocessing: 1) data collection, where medical documents are annotated by experts using ATLAS.ti for qualitative analysis; 2) positive bias filter, where potential positive biases are identified and categorized; 3) negative filters, which categorize and filter out explicitly and implicitly biased statements. After applying positive and negative filtering, we apply stratified K-fold to maintain class distribution as well as better generalization.

The textual data undergoes several preprocessing steps and we created a couple subsets of the data for each of our training cases. Regardless of the training case, we first remove XML tags and then use regular expressions to clean and standardize the text. 

\subsubsection{Positive Samples.} 
Figure~\ref{fig:upset} displays the frequency of the six most common bias types in the dataset along with the number of samples in which two or more types are simultaneously present. This suggests that approaches that can learn from samples annotated for different bias types can be beneficial for type-specific bias detection. Table~\ref{tab:Biased_Samples} shows some positive samples for different types of biases from the data.

\subsubsection{Negative Samples.}
Overall, we have divided the negative samples into four categories: from the annotated samples, we have: Explicit Negatives (EN), Implicit Negatives (IN), and Remaining Negatives (RN); in addition, we gather a set of Extracted Negatives (XN), as explained in what follows. 

The vast majority of text was reviewed yet not annotated for bias, and we can therefore considered them to be ``non-bias'' samples. To find the extracted negatives, we first remove all annotated (``positive'') quotes from the texts. After that, we filter the samples which have at least one social identifier. Those sentences will compose the XN set.

EN are annotated quotes that are coded with `non-bias' and `*-disease'. We call them explicit because they have `non-bias' label. IN are annotated quotes that are coded with `*-disease' without being explicitly labeled `non-bias' by experts. Collectively, we refer to these two categories as `hard negatives' because they needed to be reviewed by a medical expert for classification. RN are annotated quotes that do not have been coded as `*-disease', but possess other labels such as `sex misuse', `gender misuse' or `inappropriate use of language'. Table~\ref{tab:neg_sample} exemplifies different types of negative samples and Table~\ref{tab:data_charachteristics} shows the overall statistics of our dataset.

\section{Experimental Setup}


\begin{figure*}[t]
    \centering
    \includegraphics[width=.85\linewidth]{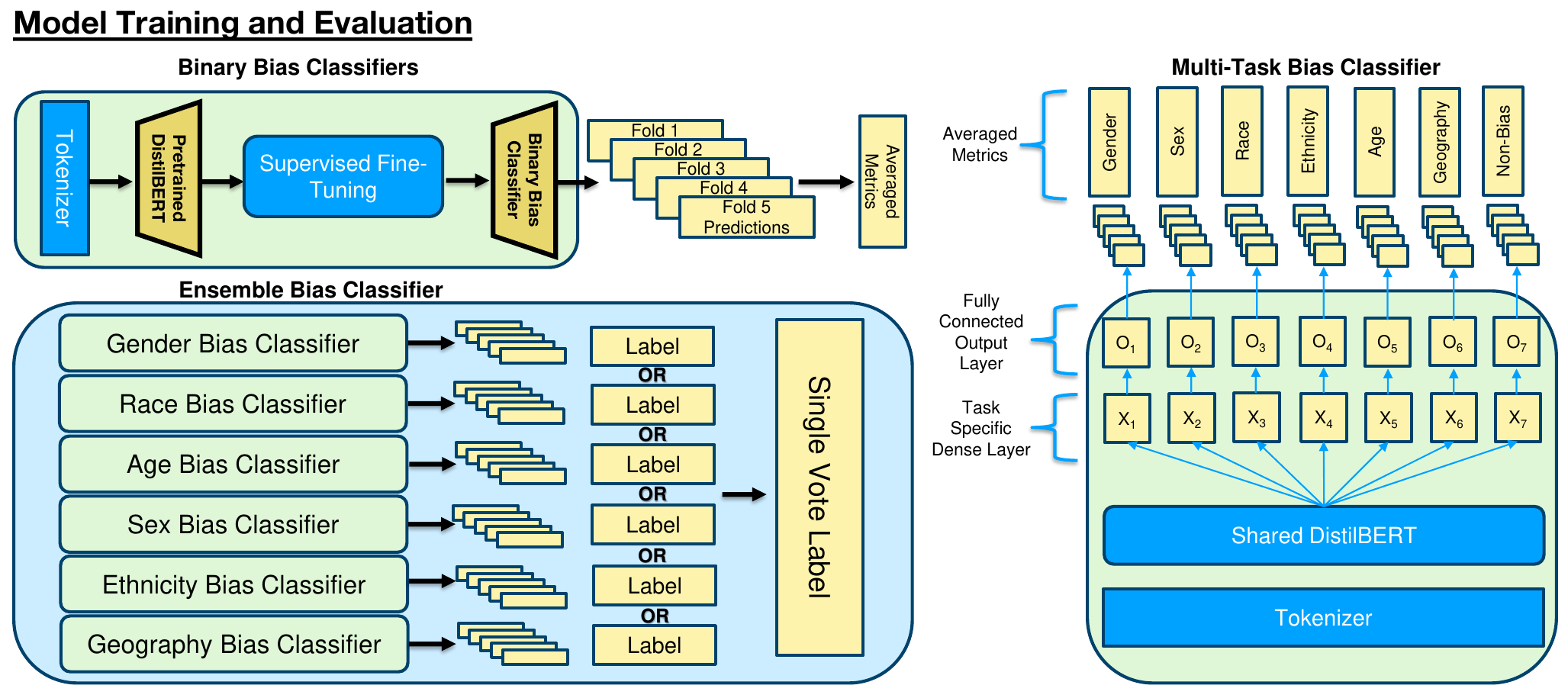}
    \caption{An overview of the model training and evaluation stage. Binary bias classifier model was used for detecting general bias as well as task-specific bias detection. The ensemble bias classifier is composed of the task-specific bias models and multitask model has 7 different tasks include bias detection as general and specific bias types. }
    \label{fig:pipeline_b}
\end{figure*}

To understand how we can construct effective machine learning (ML) models for detecting bias in medical curricular content, we investigate a few different strategies for building and training models. In this section, we present these strategies and detail our experimental setup. 

As described in Section \ref{sec:problemDef}, we refer to a total of 7 tasks:

\paragraph{Tasks 1-6: bias type-specific classification.} Given a bias type $t$, it consists of training a model $C_t$, with parameters $\theta_t$, to predict $y_t$ from some $x$ containing social identifiers associated with $t$. Predicted label is $\hat y_t = \mathds{1}\{C_t(x) > 0.5\}$.

\paragraph{Task 7: general bias classification.} Consists of training a model $C$, with parameters $\theta$, to predict $y_\mathrm{any}$ from $x$. The model output $C(x)$ is a probability, and the predicted label is $\hat y_\mathrm{any} = \mathds{1}\{C(x) > 0.5\}$, where $\mathds{1}(.)$ is an identity function.

Figure~\ref{fig:pipeline_b} illustrates our model training and evaluation, depicting the structured process of training an ensemble of binary bias classifiers and a multi-task bias classifier using a shared feature extraction layer across multiple folds, and ensemble classification. Each step integrates advanced text processing and machine learning techniques to enhance the reliability of bias detection.

\subsection{Strategies for Building Classifiers} Overall, we consider three strategies for training models, illustrated in Figure~\ref{fig:pipeline_b}:
\begin{enumerate}
\item \textbf{Binary:} In order to predict general bias, we train the model based on $y_\mathrm{any}$, without regarding specific bias types. For type-specific bias detection, we train models using $y_t$, for each bias type $t$.
\item \textbf{Ensemble:} We take the bias type-specific classifiers and combine their predictions $\hat y_t$ in a simple way, using a ``logic OR''. The rationale is that if any model trained for identifying a specific bias type predicts `bias', then the sample should be reviewed by experts. Formally, the ensemble prediction is hence defined as follows:
\[
\hat y^{\mathrm{ens}}_{\mathrm{any}}(x) = \hat y_1(x) \lor \hat y_2(x) \lor \ldots \lor \hat y_m(x).
\]

\item \textbf{Multi-task learning (MTL) classifier.} We train a single classifier capable of performing all Tasks 1-7 simultaneously via multi-task learning~\cite{goodfellow2016deep}. The goal is to enhance the performance of each task by leveraging shared knowledge across some or all of these tasks.

To do so, we build a neural network that has a shared backbone which splits into task-specific heads, each with one binary classification layer at the end. This allows the model to generate multiple outputs for the same input $x$.

Denote the function computed by the backbone by $B(.)$ and those computed by task-specific heads respectively by $H_t(.)$ for tasks \inlinemath{t \in [1..6]} and $H(.)$ for task 7. The MTL classifier outputs are calculated as
\begin{eqnarray}
     \hat y^\mathrm{mtl}_t & = & \mathds{1}\{H_t(B(x))>0.5\}, \\
     \hat y^\mathrm{mtl}_\mathrm{any} & = &\mathds{1}\{H(B(x))>0.5\},
\end{eqnarray}
for \inlinemath{t \in [1..6]}.


\textbf{Loss Function.} We choose to use a Weighted Binary Cross-Entropy as the loss function for each task to address class imbalance. This function achieves this goal by applying class-dependent weights to examples in the training data. These weights are dynamically adjusted based on the class distribution observed in the training data, ensuring that minority classes are appropriately emphasized during model training. The overall loss for the model is computed as a weighted sum of the losses from each task-specific classifier. 

\subsubsection{Weighted Binary Cross-Entropy Loss (WBCE) for a Single Task.}

Let \( y\) denote the true label for a training example and \( C(x) = P_C(y = 1 | x) \) denote the probability that the example's label is positive according to classification model $C$. The binary cross-entropy loss for this example is given by:
\[
\textsc{BCE}(y, C(x)) = - y \log(C(x)) - (1 - y) \log(1 - C(x)).
\]
When incorporating class weights, assume \( w_0 \) and \( w_1 \) are the weights for the classes 0 and 1 respectively. The weighted binary cross-entropy loss for an example $(x,y)$ becomes:
\[
\textsc{WBCE}(y, C(x)) = w_{y} \cdot \textsc{BCE}(y, C(x)).
\]
where \( w_{y} \) is selected based on the true class label \( y \) (i.e., \( w_{y} = w_0 \) if \( y = 0 \) and \( w_{y} = w_1 \) if \( y = 1 \)).

\subsubsection{Training Loss for the Multi-Task Model.}

In MTL-based frameworks, it is common to assign different weights to the loss associated with each task. However, since each input is used to generate predictions for all heads and there isn't a clear ordering of the tasks in terms of importance, we opt to assign equal weights to each.

While the inputs in the training data are the same for every task, the split between positive and negative instances differs because the positive samples for one bias type will be negative samples for another (except for multi-labeled samples). Therefore, we opt to use the task-specific class weights $w_{t,0}$ and $w_{t,1}$. Thus, the total loss used for training the MTL model is
\begin{equation}
    L = \frac{1}{N(m+1)} \sum_{t} \textsc{WBCE}(y, C(x)), 
\end{equation}
where $m+1$ is the total number of tasks.

As of May 2024, tensorflow (v2.16.1) doesn't support task-specific class weights in multiple output models. Therefore, we implemented a custom task-specific loss function and redefined \verb|tf.keras.compile| to correctly aggregate the multiple loss functions.  





\end{enumerate}

\subsection{Model Construction and Architecture}

We train several models based on DistilBERT~\cite{sanh2019distilbert}, which is a Transformer architecture based on the popular BERT model. We opt to use DistilBERT because it has been shown to achieve similar performance to BERT despite having 40\% less parameters. This largely speeds up the training process, allowing us to experiment with several configurations and perform K-fold cross-validation.

More specifically, we fine-tune the pre-trained DistilBERT model by adding classification heads, consisting of a dense layer and classification layer. This technique, which is one variant of transfer learning~\cite{goodfellow2016deep}, allows us to leverage the knowledge DistilBERT has acquired from large-scale language modeling tasks and apply it to our specific classification tasks. 
The original DistilBERT layers are frozen, while the new ones are trained by setting number of epochs to 10, learning rate to 4e-5, and batch size to 32.



Finally, we compare all our experiments to a baseline approach consisting of using FastText~\cite{bojanowski2017enriching} to obtain text embeddings and training a XGBoost~\cite{chen2016xgboost} (binary) classification model. The baseline was used to create 7 classifiers total, one for each bias type and one for the general bias task.

The training process was conducted over a maximum of 20 epochs with a dynamically calculated number of steps per epoch based on the batch size. The validation data, repeated and batched similarly to the training data, was used to evaluate the model performance at the end of each epoch.

For each setting, we set the prediction threshold to 0.5 and we report the model performances using the standard metrics of accuracy, precision, recall, F1-score, and ROC-AUC, averaged across the five folds. Since we desire to avoid false negatives, we also include F2-score, which similarly to F1 also combines precision and recall, but places more emphasis on recall. 

\subsection{Negative Training Set Variations}
\label{sec:negativeTypes}

We assess the impact of the different types of negative samples on the model's ability to differentiate biased from non-biased excerpts. To achieve this goal, we created \textbf{three variations of the training set} corresponding to different combinations of negative types. It is worth noting that the test set is kept constant across all different experiments associated with a given task.

The three variations we consider for defining the negative samples to be included the training data are as follows:
\begin{enumerate}
    \item \textbf{Extracted Negatives (XN):} we use only (negative) samples with social identifiers that were automatically extracted from our medical documents.
    \item \textbf{All Negatives (AN):} we use all types of negatives, including EN, IN, RN and XN.
    \item \textbf{All but Remaining Negatives (AN-RN):} we exclude samples from RN but consider all the others. This is due to RN containing a variety of confounding sentences with inappropriate use of language, which we suspected may reduce the models' performance.
\end{enumerate}

In contrast to most works using deep neural networks, we opt to use K-Fold cross-validation to compute the average performance based on the entire dataset. More precisely, we use a Stratified K-fold split with $K=5$ to ensures that our model is that the class distribution is consistent across the different folds.




For the binary classifier for specific bias type \inlinemath{bias(x,t)}, we used all positive bias samples for type \inlinemath{t} and concatenated those with their corresponding negative samples in each experiment. Since the ensemble model consists of the predictions made from the specific bias classifiers, they shared the same filtering protocols as their individual classifiers. For MTL, all positive cases were used with multiple labels (\inlinemath{bias(x,t)} for each type \inlinemath{t}, as well as \inlinemath{bias(x,\mathcal{T})}) and concatenated with their corresponding negative samples in each experiment. Finally, the binary classifier for the general bias detection was trained with the same sets of samples as MTL, but only using the \inlinemath{bias(x,\mathcal{T})} label.




\begin{table*}[htb!]

\caption{Performance of type-specific models for different sets of negative examples (XN: extracted negatives, AN: all negatives, AN-RN: all except remaining negatives). Best result for each pair (bias type, metric) is bold-faced.}
\label{tab:table_spec_mod}

\addtolength{\tabcolsep}{-0.05em}
\begin{tabularx}{\textwidth}{l|ccc|ccc|ccc|ccc|ccc}
    \toprule
    \textbf{Method} &\multicolumn{3}{c|}{\textbf{\underline{Precision}}} & \multicolumn{3}{c|}{\textbf{\underline{Recall}}} & \multicolumn{3}{c|}{\textbf{\underline{F1-Score}}} & \multicolumn{3}{c|}{\textbf{\underline{F2-Score}}} & \multicolumn{3}{c}{\textbf{\underline{AUC}}} \\
    {\scriptsize\textbf{(Type-specific)}} & XN & AN & {\scriptsize AN-RN} & XN & AN & {\scriptsize AN-RN} & XN & AN & {\scriptsize AN-RN} & XN & AN & {\scriptsize AN-RN} & XN & AN & {\scriptsize AN-RN} \\
    \midrule
    Gender &.292 & \textbf{.430} & .303  &\textbf{.902} & .742 &.885 & .434 & \textbf{.539} &.448 &.623 &\textbf{.643} &.633 & .948 &.948 &\textbf{.949} \\
    Sex &.340 &\textbf{.583}& .297&.839& .801 &\textbf{.868} &.480 &\textbf{.663}& .440&.643 &\textbf{.735}& .623&.932 &\textbf{.951}& .929\\
    Race &.606 &\textbf{.613}& .564&.873& .864 &\textbf{.945}&.708 &\textbf{.709}& .699&.796& .791 &\textbf{.825} &.945& .948 &\textbf{.950}\\
    Ethnicity &.615 &\textbf{.626}& .597&\textbf{.889}& .861& .852&\textbf{.725}& .722& .700&\textbf{.814}& .798& .783&\textbf{.919}& .917& .916\\
    Age &\textbf{.452}& .445& .444&.875 &\textbf{.905}& .860&.594 &\textbf{.595}& .585&.735 &\textbf{.747}& .723&.902 &\textbf{.910}& .904\\
    Geography &.522 &\textbf{.582}& .545&\textbf{.892}& .773& .849&.657& .659 &\textbf{.662}&\textbf{.780}& .721& .762&.894 &\textbf{.896} &\textbf{.896}\\
    \bottomrule
    
\end{tabularx}

\end{table*}

\begin{table}[htb!]
\caption{Performance of Binary (type-specific), Multi-Task \& Baseline (FastText+XGBoost) models on detection of each bias type, trained on all negatives (AN). Metrics: \underline{P}recision, \underline{R}ecall, \underline{F1}-score, \underline{F2}-score, \underline{AUC}. Best result for each tuple (bias type, model, metric) is bold-faced.}
\label{tab:results_an}
\begin{tabular}{clccccc}

    \toprule
   & \textbf{Method} & \textbf{\underline{P}} & \textbf{\underline{R}} & \textbf{\underline{F1}} & \textbf{\underline{F2}} & \textbf{\underline{AUC}} \\
    \midrule
     \parbox[t]{2mm}{\multirow{3}{*}{\rotatebox[origin=c]{90}{Gender}}} &Binary-Type &\textbf{.430}&\textbf{.742}&\textbf{.539}&\textbf{.643}&\textbf{.948} \\ 
    &MultiTask &.416&.406&.396&.397&.868 \\
    &Baseline &.067&.333&.111&.187&.617\\
    \midrule
    \parbox[t]{2mm}{\multirow{3}{*}{\rotatebox[origin=c]{90}{Sex}}} &Binary-Type &.583&\textbf{.801}&\textbf{.663}&\textbf{.735}&\textbf{.951} \\ 
    &MultiTask &\textbf{.632}&.478&.531&.496&.886 \\
    &Baseline &.097&.199&.176&.188&.576\\

    \midrule
    \parbox[t]{2mm}{\multirow{3}{*}{\rotatebox[origin=c]{90}{Race}}} &Binary-Type &.613&\textbf{.864}&.709&\textbf{.791}&\textbf{.948} \\ 
    &MultiTask &\textbf{.764}&.756&\textbf{.753}&.753&\textbf{.948} \\
    &Baseline &.018&.324&.034&.074&.548\\

    \midrule
    \parbox[t]{2mm}{\multirow{3}{*}{\small\rotatebox[origin=c]{90}{Ethnicity}}} &Binary-Type &.626&\textbf{.861}&\textbf{.722}&\textbf{.798}&\textbf{.917} \\ 
    &MultiTask &\textbf{.712}&.695&.695&.693&.908\\
    &Baseline &.020&.370&.038&.082&.568\\

    \midrule
    \parbox[t]{2mm}{\multirow{3}{*}{\rotatebox[origin=c]{90}{Age}}} &Binary-Type &.445&\textbf{.905}&\textbf{.595}&\textbf{.747}&\textbf{.910} \\ 
    &MultiTask &\textbf{.580}&.383&.445&.403&.839 \\
    &Baseline &.117&.428&.184&.280&.634\\

    \midrule
    \parbox[t]{2mm}{\multirow{3}{*}{\scriptsize \rotatebox[origin=c]{90}{Geography}}} &Binary-Type &.582&\textbf{.773}&\textbf{.659}&\textbf{.721}&\textbf{.896} \\ 
    &MultiTask &\textbf{.693}&.615&.645&.625&.887 \\
    &Baseline &.041&.538&.076&.158&.651\\
    \bottomrule
    
\end{tabular}

\end{table}

\begin{table*}[htb!]
\caption{Performance of bias detection for different sets of negative examples (XN: extracted negatives, AN: all negatives, AN-RN: all except remaining negatives). Best result for each pair (bias type, metric) is bold-faced. Overall best for each metric is underlined. AUC is not applicable to the Ensemble method, since it is generated using a logical OR between binary outputs.}
\label{tab:general_bias_classifiers}
\addtolength{\tabcolsep}{-0.1em}
\begin{tabularx}{\textwidth}{l|ccc|ccc|ccc|ccc|ccc}
    \toprule
    \textbf{Method} &\multicolumn{3}{c|}{\textbf{\underline{Precision}}} & \multicolumn{3}{c|}{\textbf{\underline{Recall}}} & \multicolumn{3}{c|}{\textbf{\underline{F1-Score}}} & \multicolumn{3}{c|}{\textbf{\underline{F2-Score}}} & \multicolumn{3}{c}{\textbf{\underline{AUC}}} \\
    {\scriptsize} & XN & AN & {\scriptsize AN-RN} & XN & AN & {\scriptsize AN-RN} & XN & AN & {\scriptsize AN-RN} & XN & AN & {\scriptsize AN-RN} & XN & AN & {\scriptsize AN-RN} \\
    \midrule
    {\small Binary-General} &.314 & \textbf{.504} & .335  &\textbf{.925} & .812 &.893 & .468 & \underline{\textbf{.615}} &.486 &.665 &\underline{\textbf{.717}} &.668 & .880 &\underline{\textbf{.923}} &.887 \\
    Ensemble &.206 &\textbf{.232}& .219&\underline{\textbf{.954}}& .936 &.951 &.338 &\textbf{.371}& .356&.552 &\textbf{.580}& .570&-&-&-\\
    MultiTask &.400 &\underline{\textbf{.624}}& .438&\textbf{.780}& .479 &.724&.527 &.530& \textbf{.541}&\textbf{.654}& .495 &.636 &\textbf{.874}& .869 &\textbf{.874}\\
    Baseline &\textbf{.288} &.263&.260 &.461&.454 &\textbf{.527}&\textbf{.354}&.333 &.348 &.411&.396 & \textbf{.437}&.708&.709&\textbf{.722} \\
    \bottomrule
\end{tabularx}
\label{tab:tab_general}
\end{table*}

\begin{figure}[tb!]
    \centering
    \begin{subfigure}{.7\columnwidth}
        \includegraphics[width=\linewidth]{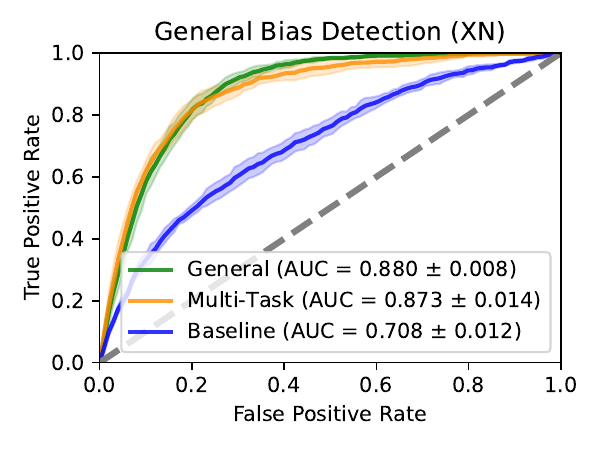}
        \label{fig:auc-e1}
    \end{subfigure}
    \begin{subfigure}{.7\columnwidth}
        \includegraphics[width=\linewidth]{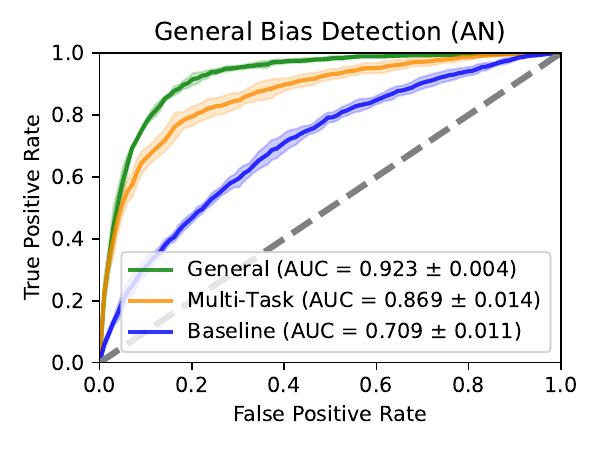}
        \label{fig:auc-e2}
    \end{subfigure}
    \begin{subfigure}{.7\columnwidth}
        \includegraphics[width=\linewidth]{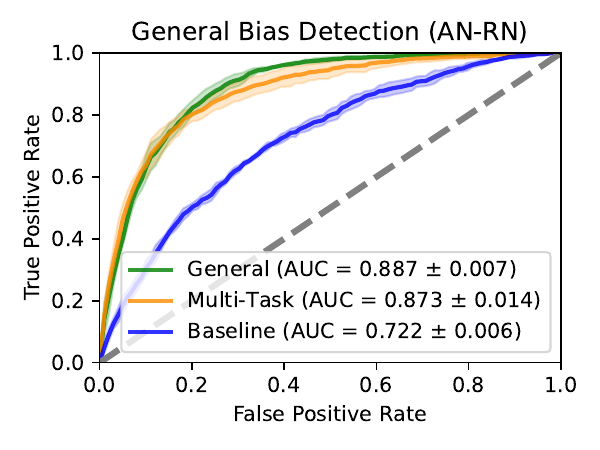}
        \label{fig:auc-e2}
    \end{subfigure}

\caption{Mean ROC curves and standard deviation for general bias classification when training with different sets of negative samples (Top: XN, Middle: AN; Bottom: AN-RN).}
\label{fig:AUC_general}
\end{figure}

\section{Results}

 In this section, we present the results of our proposed type-specific, general, ensemble and multi-task learning models and compare their performance with a baseline model in the context of specific and general bias detection tasks. Through a detailed analysis employing ROC curves and AUC metrics, we evaluate the effectiveness of our models across different folds of validation, thereby ensuring robustness and consistency in our evaluations.
 
 Firstly, we discuss the performance of models in the detection of specific bias types. The first comparison focuses on the binary classification models. Table~\ref{tab:table_spec_mod} shows their performance across all of our experiment settings. We found that training with only the XN negative category leads to the high recall, at the detriment of lower precision (except for type `ethnicity'), ultimately leading to low F1-score. Between training with AN and AN-RN, we find the results to be generally comparable, however, AN usually yields AUC that is higher or comparable to that achieved by AN-RN.

 Based on the previous observation, we choose the setting AN to compare the binary classifier with the Multi-Task and baseline methods. Table~\ref{tab:results_an} presents the results of the comparison. Both the type-specific and multitask model greatly outperformed the baseline model. While MTL often led to the best precision, the binary classifier has outperformed MTL in all the other metrics. Furthermore, the AUC for the latter model across different tasks varied from 0.896 to 0.951, which is considered very effective at detecting bias.
 
 Last, we compare the proposed models and the baseline on the general bias classification task.
 Table~\ref{tab:general_bias_classifiers} shows the performance results for the different experiment settings and metrics. Once again, the baseline did not perform well. Considering only the proposed methods, the AN setting generally led better precision, whereas the XN setting led to better recall. Not surprisingly, the Ensemble method achieves the best overall recall, since the logical OR will cause the model to be highly sensitive, leading to the lowest precision in our results. Based on the F1-Score, F2-Score and AUC, the best result is achieved by the binary classifier model with the AN setting. This result is consistent with that for type-specific bias classification. For other settings (i.e., XN and AN-RN), MTL yields similar F2-Score and AUC to the binary classifier.
 
 

 We include ROC curves for the baseline and the proposed methods (except for the Ensemble method, since their predictions are based on the logical OR and cannot be directly converted to probabilities) in Figure~\ref{fig:AUC_general}. These figures highlight the superiority of the binary classifier, particularly with the AN setting. Based on Figure~\ref{fig:AUC_general}(middle), we can see that this combination would allow one to choose a threshold that achieves a high true positive rate without incurring in high false positive rates.

\section{Conclusion and Future Work}
Our investigation into bisinformation in medical education has revealed a significant issue that, if left unaddressed, can continue to negatively impact health quality, equity, and outcomes. Through a detailed empirical study, we have made substantial progress in identifying and mitigating bias in medical education materials related to gender, sex, age, race, and ethnicity. 

Our work represents the first comprehensive attempt to remove ingrained biases from medical curricula using a systematic and scalable AI-based approach. This paper provides numerous contributions: not only a robust dataset and bias detection classifier but also a pioneering methodology that combines modern AI techniques with a deep understanding of medical educational material. Our proposed method has demonstrated a significant capability in distinguishing biased content, as confirmed by our performance evaluations.

Moreover, our research extends beyond academic boundaries and ventures into the realm of public health and online communities. By adapting our classifier to comprehend and interpret the subtleties of social media platforms like X (formerly Twitter), we strive to bridge the gap between conventional medical instruction and contemporary digital communication channels. This is of utmost importance, especially in the face of current public health challenges and the rampant dissemination of bisinformation.

The present paper lays the groundwork for eliminating biases from medical curricula and sets the stage for a paradigm shift in medical education. In this context, our models 
make the first strides towards leveraging artificial intelligence to promote fairness and integrity in medical education by addressing biases that disproportionately affect vulnerable communities, particularly those underserved and often misrepresented in medical literature, has been pivotal. By advocating for those most impacted by biases, we ensure that the outcomes of our work are both socially informed and ethically sound.

Our future work will delve into visual content, extending our analytical lens to the biases inherent in images within medical curricula. We recognize that imagery can convey powerful and often subconscious messages. We aim to gain a deep understanding of how visual representations contribute to the perpetuation of bias. Additionally, we will investigate other potential biases in medical texts, such as those related to socioeconomic status, disability, neurodiversity, and intersectionality, to further enhance our understanding of bias's multifaceted and complex nature in medical education.

Lastly, another critical avenue for our forthcoming research is examining how biased data proliferates through online social media platforms. The velocity and volume with which information circulates in the digital space can amplify the dissemination of biased content, leading to widespread misconceptions and reinforcing harmful stereotypes. We plan to employ and enhance our models to track and analyze the trajectory of such misinformation on social media. Through this, we aim to contribute to developing more robust countermeasures against the spread of medical misinformation, ultimately helping safeguard public health discourse in the digital age.

\section*{Ethics Statement}

We include the following information for context on the ethics of our project.

\paragraph{Ethical Considerations.}

This study is rooted in making strides toward facilitating institutional change in the medical field. We are developing this software to be used with human intervention, requiring a final stage of a human review. These algorithms should not decide whether specific material contains biased information but rather be used as a tool for educators to be aware of possible bias within their curricula. It is up to each institution to ensure that we move beyond identifying bias to implementing strategies to reduce bias in curricular content. Likewise, institutions should be held responsible for ensuring their faculty work on minimizing thier own biases to avoid perpetuating bisinforamation that can potentially harm patients and their health outcomes. It is up to medical institutions to provide appropriate faculty development and resources to accomplish this large undertaking. Lastly, it's important not to burdened our under-represented faculty with this task given and inadvertently adding to the minority tax burden.\\


\paragraph{Adverse Impacts}

This work should never be taken as a way to penalize educators but rather a tool for fostering growth - to identify bias and allow the opportunity to correct it. When implementing this approach in practical scenarios, it is crucial to consider the context of the findings and not automatically assume that correlation implies causation. For instance, discovering higher levels of potential bias in a school's curriculum does not necessarily mean that the faculty or students are inherently more biased. \\

\paragraph{Conflicts of Interest.} At the time this project was initiated, [name redacted] held a significant financial interest in [company name redacted].

\paragraph{Researcher Positionality.}

The primary investigators and trainees have multiple identities that have diverse backgrounds and abilities. Participants in this work stem from multiple backgrounds including computer science, social science and medicine. These identities have influenced our specific lens on these issues and as such we acknowledge that there are both benefits and limitations stemming from our intersecting identities. 

The authors note that ChatGPT was used for light editing only. The authors remain responsible for all content.

\bibliography{refs}

\appendix

\section{Appendix: Coding Guidelines}

The BRICC Project created a comprehensive Coding Guideline using social science principles to annotate medical text systematically. In what follows, we provide two brief snapshots from the 20-page guideline.

\begin{figure*}[]
    \centering
    \frame{\makebox[\textwidth]{\includegraphics[width=.9\paperwidth]{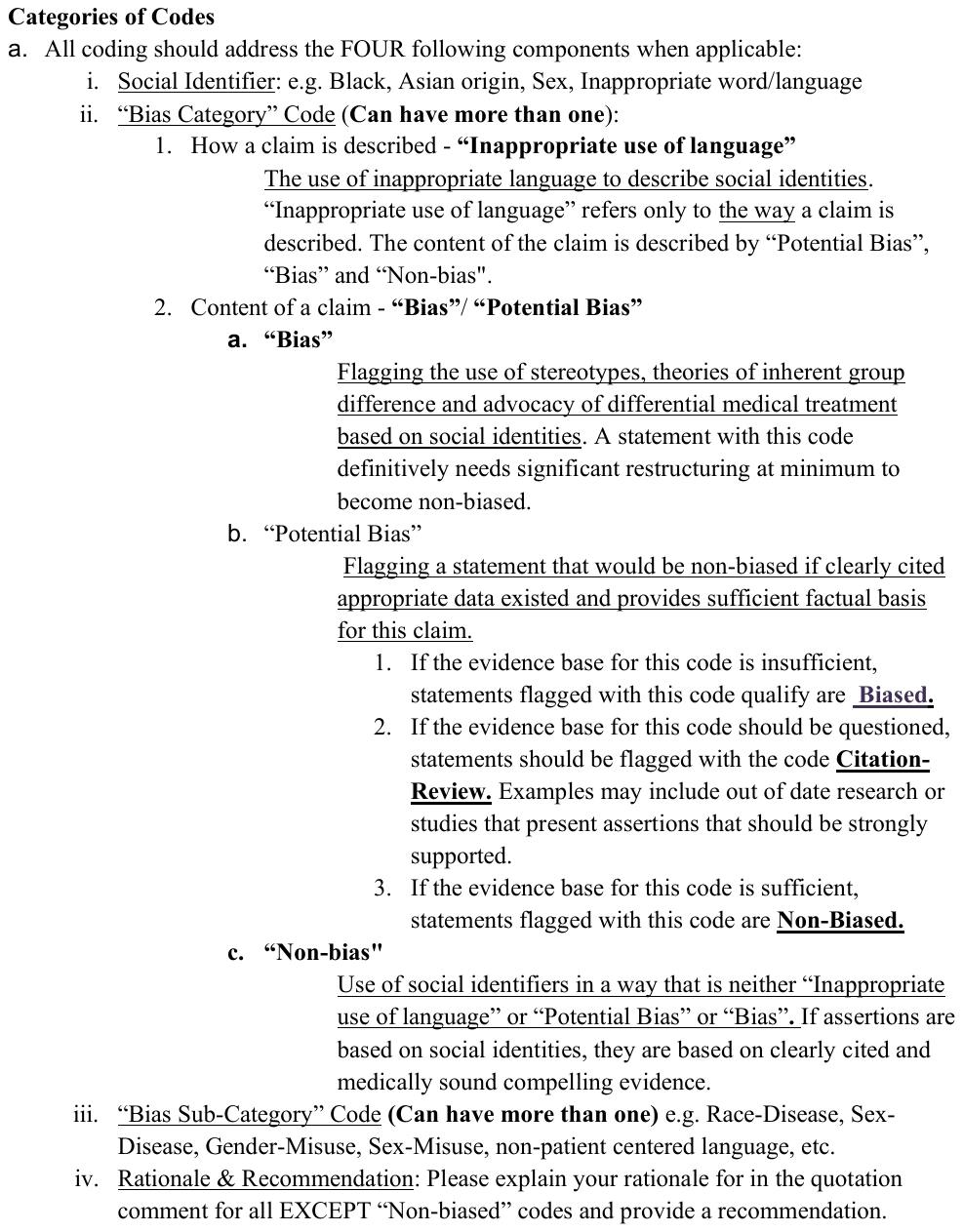}}}
    \caption{Introductory section of the BRICC Coding Guidelines, which are generic and apply to all types of Bias.}
    \label{fig:guideline}
\end{figure*}

\begin{figure*}[]
    \centering
    \frame{\makebox[\textwidth]{\includegraphics[width=.9\paperwidth]{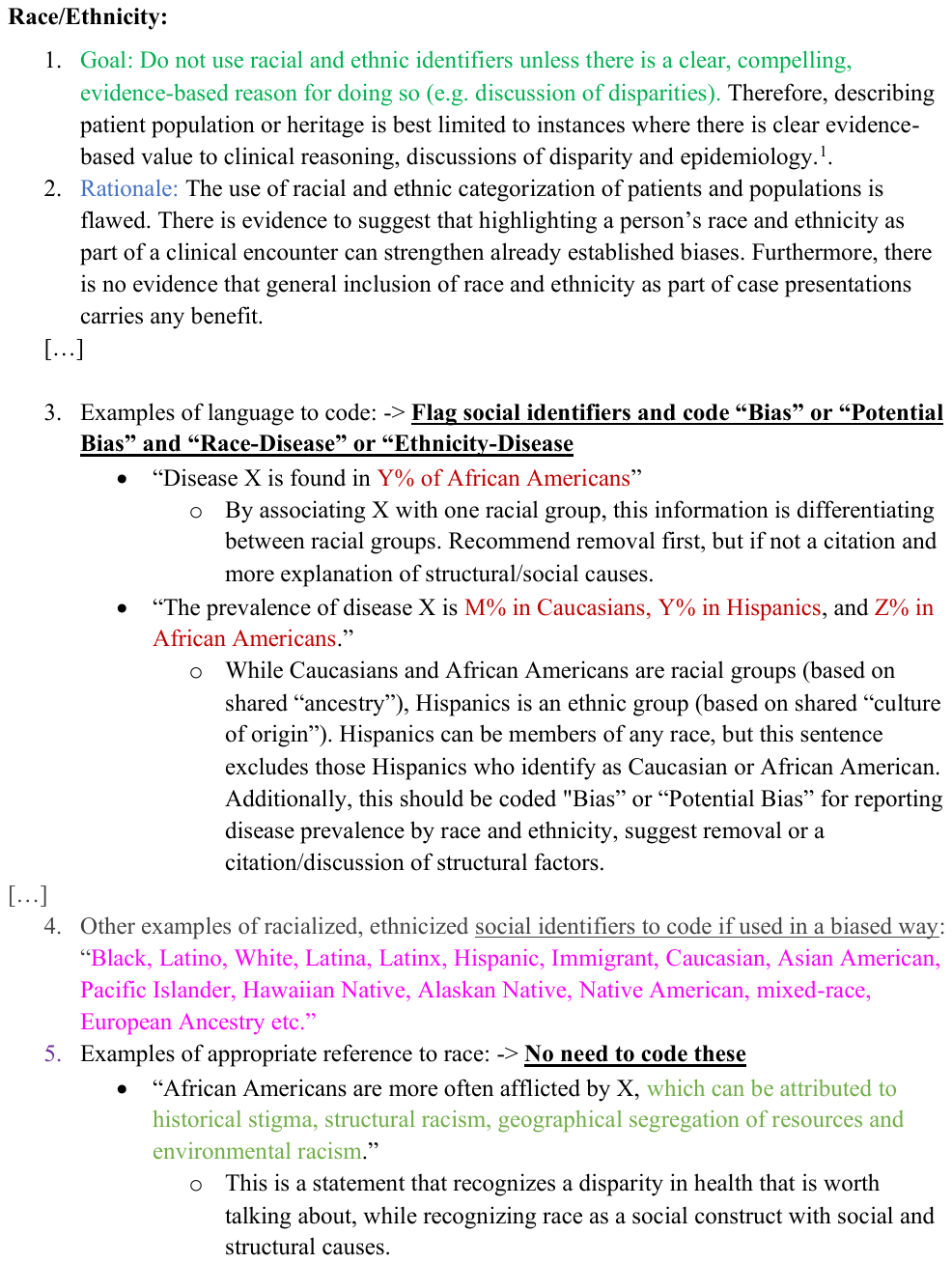}}}
    \caption{A snapshot from the BRICC Coding Guidelines under the Race/Ethnicity category. [...] denotes omission for brevity reasons.}
    \label{fig:guideline}
\end{figure*}

\end{document}